\documentclass[reprint,superscriptaddress,aps,twocolumn,showpacs]{revtex4-2}
\usepackage{amssymb}
\usepackage{amsmath}
\usepackage{graphicx}
\usepackage{multirow}
\usepackage{subcaption}
\allowdisplaybreaks
\usepackage{physics}
\usepackage{hyperref}
\usepackage{makecell}

\begin{document}

%\preprint{APS/123-QED}

\title{$P_c$ states in the mixture of molecular and pentaquark pictures}

\author{K. Xu}
\email{gxukai1123@gmail.com}
\author{K. Phumphan}
\email{k.phoompan393@hotmail.com}
\author{W. Ruangyoo}
\author{C. C. Chen}
\address{School of Physics and Center of Excellence in High Energy Physics and Astrophysics, Suranaree University of Technology, Nakhon Ratchasima 30000, Thailand}
\address{Department of Physics, National Cheng Kung University, Tainan, 70101, Taiwan}
\author{A. Limphirat}
\email{ayut@g.sut.ac.th}
\author{Y. Yan}
\email{yupeng@g.sut.ac.th}
\address{School of Physics and Center of Excellence in High Energy Physics and Astrophysics, Suranaree University of Technology, Nakhon Ratchasima 30000, Thailand}
\date{\today}% It is always \today, today,

\begin{abstract}
We systematically study hidden charm pentaquark states in the constituent quark model with a general Hamiltonian for multiquark systems, considering the coupling between  
the $\Sigma_c^{(*)}\bar{D}^{(*)}$ molecular states and the $q^3c\bar c$ compact pentaquark states by the one-gluon exchange hyperfine interaction. The ground state hidden-charm pentaquark mass spectra and the strong decay widths are calculated. This work suggests that $P_c(4312)$, $P_c(4457)$ and $P_c(4380)$ resonances might be mainly $\Sigma_c \bar D$, $\Sigma_c \bar D^*$ and $\Sigma_c^* \bar D$ hadronic molecules respectively, and $P_c(4440)$ might include sizable pentaquark components.

 %
%The $P_c$ compact pentaquark states, if exist, should lie in the higher regions than the observed states like, 4480, 4510 and 4529 MeV and decay largely in the open-charm channels.  considering the coupling of the .

\end{abstract}

%\begin{keyword}
%Mixture of molecular and compact pentaquarks, Total and partial decay width 
%\end{keyword}

\maketitle

%\tableofcontents
\section{\label{sec1}Introduction}

The pentaquark-like states, $P_c(4380)$, $P_c(4312)$, $P_c(4440)$, and $P_c(4457)$ have been studied widely in both the hadronic molecular and compact pentaquark scenarios since their discovery by LHCb \cite{LHCb1,LHCb2,LHCb3}. One may see some good reviews in Refs. \cite{chenreview2016,Espositoreview2017,Alireview2017,Guoreview2018,Olsen2018,sym15071298}. As the masses of the $P_c$ states are several MeV below the $\Sigma_C \bar D$ and $\Sigma_c \bar D^*$ mass thresholds, the $P_c$ states have been interpreted as hadronic molecules in various approaches such as QCD sum rules \cite{Chen_2015, Chen:2019bip, Azizi_2017, Azizi_2018, Azizi_2018dva,Wang:2019got}, potential models \cite{Guo2019,Mut2019,Zhu2019,Eides2019,Weng2019,Wang2020}, effective field theory \cite{Liu2019,He2019}, heavy hadron chiral perturbation theory \cite{Meng2019}, heavy quark spin multiplet structures \cite{Xiao22019}, phenomenological Lagrangian approach \cite{Thomas2019} and constituent quark models \cite{Huang2019}.
In the compact pentaquark picture, the $P_c$ states have been studied as diquark-diquark-antiquark structures in the constituent quark-diquark model \cite{Ali2019theo}. The $P_c$ mass spectrum and decay widths are studied in a potential model \cite{Gian2019}, QCD sum rules \cite{Wang2020}, chromomagnetic (CM) model \cite{Zhupenta2019}, and non-relativistic constituent quark model \cite{RZhu2019}.
The $P_c$ states have also been studied in the mixing scenario of the baryon-meson molecule and compact pentaquark states \cite{He2005,Yama2017,Yama2021}.  The masses and widths of the $P_c$ resonances are explained by coupling the $\Sigma^{(*)}_c\bar{D}^{(*)}$ hadronic molecule with a compact five-quark state in Refs. \cite{Yama2017,Yama2021}. 
%\cite{Ali2019theo,Gian2019,Wang2020} 

In the present work, we extend the non-relativisitic constitute quark model, where a Cornell-like potential with one-gluon hyperfine interaction is employed~\cite{Kai2020PRD}, to study the ground state $q^3c\bar c$ pentaquark mass spectrum, the coupling between the compact pentaquark states and charmed baryon-meson molecules, and decays of the coupled states. It is found that one-gluon exchange hyperfine interactions not only mix up compact pentaquark states of the different configurations, but also couple the hadronic molecules and the compact pentaquark states of the same quantum numbers.

The paper is arranged as follows. In Sec.~\ref{sec2}, we briefly introduce the Hamiltonian for multi-quark systems. The $q^3c\bar c$ pentaquark mass spectra are evaluated in the mixing picture of molecular and pentaquark states and presented in Sec.~\ref{sec2}. In Sec.~\ref{sec4} we calculate the partial strong decay widths of the $P_c$ states. Discussion and summary are given in Sec.~\ref{sec4}. Sec.~\ref{sec5}.

\section{\label{sec2}THEORETICAL MODEL and pentaquark mass spectrum}
We start with the general Hamiltonian \cite{Kai2019PRC,Kai2020PRD,Zhao2021PRD},
\begin{flalign}\label{eqn::ham}
	&H =H_0+ H_{hyp}^{OGE}, \nonumber \\
	&H_{0} =\sum_{k=1}^{N} (m_k+\frac{p_k^2}{2m_{k}})+\sum_{i<j}^{N}(-\frac{3}{8}\lambda^{C}_{i}\cdot\lambda^{C}_{j})(A_{ij} r_{ij}-\frac{B_{ij}}{r_{ij}}),  \nonumber \\
	&H_{hyp}^{OGE} = -\frac{C_m}{m_u^2}\sum_{i<j}\frac{\lambda^{C}_{i}\cdot\lambda^{C}_{j}}{m_{i}m_{j}}\,\vec\sigma_{i}\cdot\vec\sigma_{j},
\end{flalign}
where $\lambda^C_{i}$ are the generators of color SU(3) group, and $A_{ij}$ and $B_{ij}$ are mass-dependent coupling constants, taking the form, 
\begin{eqnarray}
	A_{ij}= a \sqrt{\frac{m_{ij}}{m_u}},\;\;B_{ij}=b \sqrt{\frac{m_u}{m_{ij}}}.
\end{eqnarray}
with $m_{ij}$ being the reduced mass of the $i$th and $j$th quarks, defined as $m_{ij}=\frac{2m_i m_j}{m_i+m_j}$.
The hyperfine interaction $H_{hyp}^{OGE}$, including only the one-gluon exchange contribution, couples together the hadronic molecules and compact pentaquark states. 
In the study, the model parameters of the three coupling constants and four constituent quark masses, taken directly from the previous works \cite{Kai2019PRC,Kai2020PRD}, are determined
by fitting the theoretical results to the mass of the ground state baryons and mesons. They are
\begin{eqnarray}\label{eq:nmo}
	&
	m_u = m_d = 327 \ {\rm MeV}\,, \quad
	m_s = 498 \ {\rm MeV}\,, \nonumber\\
	&
	m_c = 1642 \ {\rm MeV}\,, \quad
	m_b = 4960 \ {\rm MeV}\,, \nonumber\\
	&
	C_m   =  18.3 \ {\rm MeV}, \quad
	a     = 49500 \ {\rm MeV^2}, \quad
	b     =  0.75 \nonumber\\
\end{eqnarray}
We consider the coupling of the $S$-wave molecular states, $\Sigma^{*}_c\bar{D}^{*}$, $\Sigma_c\bar{D}^{*}$, $\Sigma_c\bar{D}$ with the ground hidden-charm pentaquarks. The $\Lambda_c^{+}\bar D^{(*)0}$ is excluded since the $\Lambda_c^{+}$ and $\bar D^{(*)0}$ interaction is likely repulsive \cite{wang2011}.  By solving the coupled Schr{\"o}dinger equations, we derive the eigenstates of the $\Sigma^{*}_c\bar{D}^{*}$, $\Sigma_c\bar{D}^{*}$, $\Sigma_c\bar{D}$ systems plus compact pentaquark states. The results are presented in Tables \ref{tab:mol1} and \ref{tab:i32} for isospin 1/2 and 3/2 respectively, where $M$ stands for the eigenvalues and $|A_i|^2$  for the contribution of all coupled configurations. In the calculations, the wave functions of compact $q^3c\bar c$ pentaquark states, $\Sigma^{*}_c$, and $\bar{D}^{*}$ are directly taken from Refs. \cite{Kai2020PRD,Zhao2021PRD}. And we have applied the approximation that the mass of the $S$-wave molecular components, $\Sigma^{*}_c\bar{D}^{*}$, $\Sigma_c\bar{D}^{*}$, $\Sigma_c\bar{D}$ is set to the mass threshold.

\begin{table}[b]
\caption{\label{tab:mol1}Mixtures of all hadronic molecules and compact pentaquark states for spin 3/2 and 1/2 of $I=1/2$. All mass units are in MeV.}
\begin{ruledtabular}
\begin{tabular}{@{}cccc}
$J^P$ &  Mixing states & $|A_{i}|^2$ & Mass  \\ [2pt]
      \hline 
      $\frac{1}{2}^-$&$\left\lbrace \begin{array}{c} \Sigma_c^{*}\overline D^{*} (4526)\\   \Psi^{csf}_{[21]_{C}[21]_{F}[21]_{S}} \\ \Psi^{csf}_{[21]_{C}[21]_{F}[21]_{S}} \\ \Psi^{csf}_{[21]_{C}[21]_{F}[3]_{S}}\end{array} \right\rbrace $  & $\left[
      \begin{array}{cccc}
0.50&0.18&0.04&0.28\\
0.47&0.33& &0.21\\
0.03&0.18&0.71&0.08\\
 &0.31&0.26&0.43\\ 
\end{array}
\right]$  & $\left( \begin{array}{c} 4535 \\4517 \\4455 \\ 4433 \end{array} \right)$ \\[2pt]
   $\phantom{-}$& $\left\lbrace \begin{array}{c} \Sigma_c\overline{D}^{*}(4462) \\  \Psi^{csf}_{[21]_{C}[21]_{F}[21]_{S}} \\ \Psi^{csf}_{[21]_{C}[21]_{F}[21]_{S}} \\ \Psi^{csf}_{[21]_{C}[21]_{F}[3]_{S}}\end{array} \right\rbrace $  & $\left[
\begin{array}{cccc}
 &0.48&0.02&0.50\\
0.55&0.04&0.42& \\
0.22&0.38&0.10&0.30\\
0.24&0.10&0.47&0.20\\
\end{array}
\right]$  & $\left( \begin{array}{c} 4526 \\4479 \\ 4444 \\4426 \end{array} \right)$ \\[2pt]
   $\phantom{-}$& $\left\lbrace \begin{array}{c} \Sigma_c\overline{D} (4322)\\  \Psi^{csf}_{[21]_{C}[21]_{F}[21]_{S}} \\ \Psi^{csf}_{[21]_{C}[21]_{F}[21]_{S}} \\ \Psi^{csf}_{[21]_{C}[21]_{F}[3]_{S}}\end{array} \right\rbrace $  & $\left[
\begin{array}{cccc}
 &0.49&0.02&0.49\\
0.03&0.38&0.34&0.25\\
0.09&0.09&0.61&0.21\\
0.88&0.05&0.02&0.06\\
\end{array}
\right]$  & $\left( \begin{array}{c} 4526 \\4458 \\ 4451 \\4298 \end{array} \right)$ \\[2pt]   
   \hline
      $\frac{3}{2}^-$ & $\left\lbrace \begin{array}{c} \Sigma_c^{*}$$\overline D^{*} (4526)\\ \Psi^{csf}_{[21]_{C}[21]_{F}[21]_{S}} \\ \Psi^{csf}_{[21]_{C}[21]_{F}[3]_{S}} \\ \Psi^{csf}_{[21]_{C}[21]_{F}[3]_{S}}\end{array} \right\rbrace $ & $\left[
\begin{array}{cccc}
0.20&0.12&0.64&0.04\\
 &0.08&0.11&0.81\\
0.77&0.10&0.13& \\
0.02&0.70&0.13&0.16\\
\end{array}
\right]$  & $\left( \begin{array}{c} 4586 \\4532 \\4509 \\ 4473 \end{array} \right)$ \\[2pt]
      $\phantom{-}$ &$\left\lbrace \begin{array}{c} \Sigma_c^{*}\overline{D}(4386) \\ \Psi^{csf}_{[21]_{C}[21]_{F}[21]_{S}} \\ \Psi^{csf}_{[21]_{C}[21]_{F}[3]_{S}} \\ \Psi^{csf}_{[21]_{C}[21]_{F}[3]_{S}}\end{array} \right\rbrace $ & $\left[
\begin{array}{cccc}
&0.18&0.77&0.05\\
 &0.08&0.13&0.79\\
0.05&0.69&0.11&0.16\\
0.95&0.05& & \\
\end{array}
\right]$  & $\left( \begin{array}{c} 4571 \\4532 \\ 4479 \\4376 \end{array} \right)$ \\[2pt]
      $\phantom{-}$ &$\left\lbrace \begin{array}{c} \Sigma_c\overline D^{*} (4462)\\ \Psi^{csf}_{[21]_{C}[21]_{F}[21]_{S}} \\ \Psi^{csf}_{[21]_{C}[21]_{F}[3]_{S}} \\ \Psi^{csf}_{[21]_{C}[21]_{F}[3]_{S}}\end{array} \right\rbrace $ & $\left[
\begin{array}{cccc}
  &0.17&0.78&0.04\\
0.01&0.08&0.11&0.80\\
0.03&0.73&0.11&0.13\\
0.95&0.02& &0.02\\
\end{array}
\right]$  & $\left( \begin{array}{c} 4570 \\4533 \\ 4474 \\4457 \end{array} \right)$ \\[2pt]
\end{tabular}
\end{ruledtabular}
\end{table}

 \begin{table}[tb]
\caption{Mixtures of hadronic molecules and compact pentaquark states for I= 3/2.}\label{tab:i32}
\begin{ruledtabular}
\begin{tabular}{@{}cccc}
$J^P$ &  Mixing states & $|A_{i}|^2$ & Mass  \\ [2pt]
   \hline 
 $\frac{1}{2}^-$& $\left\lbrace \begin{array}{c} \Sigma_c^* \overline D^* (4526)\\ \Psi^{csf}_{[21]_{C}[3]_{F}[21]_{S}} \\ \Psi^{csf}_{[21]_{C}[3]_{F}[21]_{S}}\end{array} \right\rbrace $  & $\left[
\begin{array}{ccc}
  0.20 & 0.47 & 0.33 \\
 0.01 & 0.37 & 0.62 \\
 0.80 & 0.15 & 0.05 \\
\end{array}
\right]$& $\left( \begin{array}{c} 4813 \\ 4661 \\ 4452\end{array} \right)$ \\ [2pt]
    $\phantom{-}$ &  $\left\lbrace \begin{array}{c} \Sigma_c \overline D^* (4462) \\ \Psi^{csf}_{[21]_{C}[3]_{F}[21]_{S}} \\ \Psi^{csf}_{[21]_{C}[3]_{F}[21]_{S}}\end{array} \right\rbrace$  & $\left[
\begin{array}{ccc}
 0.05 & 0.31 & 0.64 \\
 0.08 & 0.68 & 0.24 \\
  0.88 & 0.01 & 0.11 \\
\end{array}
\right]$& $\left(\begin{array}{c}  4755\\4683 \\4423 \end{array} \right)$ \\ [2pt]
       $\phantom{-}$ &  $\left\lbrace \begin{array}{c} \Sigma_c \overline D (4322) \\ \Psi^{csf}_{[21]_{C}[3]_{F}[21]_{S}} \\ \Psi^{csf}_{[21]_{C}[3]_{F}[21]_{S}}\end{array} \right\rbrace$ &   $\left[
\begin{array}{ccc}
 & 0.52 & 0.48 \\
 0.02 & 0.47 & 0.52 \\
  0.98 & 0.01 & 0.01 \\
\end{array}
\right]$& $\left( \begin{array}{c}  4744\\4665 \\4311 \end{array} \right)$ \\ [2pt]
   \hline
    $\frac{3}{2}^-$ &  $\left\lbrace \begin{array}{c} \Sigma_c^* \overline D^*(4526) \\ \Psi^{csf}_{[21]_{C}[3]_{F}[21]_{S}}\end{array} \right\rbrace $  & $\left[
\begin{array}{cc}
 0.16 & 0.84 \\
 0.84 & 0.16 \\
\end{array}
\right]$ & $\left( \begin{array}{c}  4745 \\4480 \end{array} \right)$ \\[2pt]
     $\phantom{-}$ &  $\left\lbrace \begin{array}{c} \Sigma_c \overline D^* (4462)\\ \Psi^{csf}_{[21]_{C}[3]_{F}[21]_{S}}\end{array} \right\rbrace $ & $\left[
\begin{array}{cc}
 0.02 & 0.98 \\
  0.98 & 0.02 \\
\end{array} \right]$ & $\left( \begin{array}{c}  4706\\  4454 \end{array} \right)$ \\[2pt]
    $\phantom{-}$ &  $\left\lbrace \begin{array}{c} \Sigma_c^* \overline D (4386)\\ \Psi^{csf}_{[21]_{C}[3]_{F}[21]_{S}}\end{array} \right\rbrace $  & $\left[
\begin{array}{cc}
 0.04 & 0.96 \\
   0.96 & 0.04 \\
\end{array}\right]$ & $\left( \begin{array}{c} 4714 \\ 4369 \end{array} \right)$ \\[2pt]
\end{tabular}
\end{ruledtabular}
\end{table}

\begin{figure}[b]
	\centering
	\includegraphics[width=0.45\textwidth]{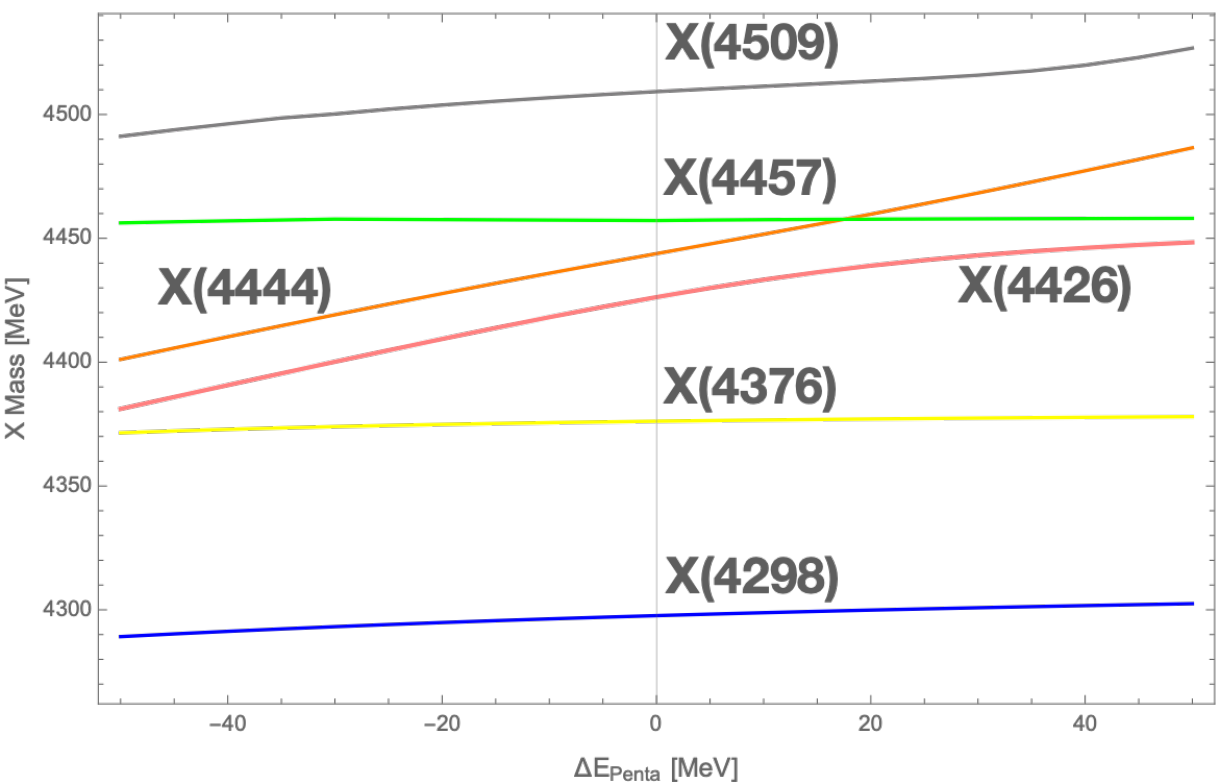}
%	\vspace{-1\baselineskip}
	\caption{X mass dependence on the the mass of pentaquark components}\label{runmass}
\end{figure}

  \begin{table*}[t]
  \setlength{\tabcolsep}{2pt}
 \centering
 \caption{The relative partial decay width of bound states X of isospin 1/2 which are normalized to the total decay width of X(4457). All mass units are in MeV.}\label{molepartial1}
 \begin{ruledtabular}
 \begin{tabular}{ccccccccccccc}
 $J$&Threshold& Mass  &$\mathrm{Eigenvector}^2$&Total&$p\eta_c$&$pJ/\psi$&$\Sigma_c^*\bar{D}$&$\Sigma_c\bar{D}$&$\Lambda_c^+\bar{D}$&$\Sigma_c^*\bar{D}^*$&$\Sigma_c\bar{D}^*$&$\Lambda_c^+\bar{D}^*$\\ [2pt]
 \hline
 $\frac{1}{2}$ & $\Sigma_c\bar{D}(4322)$&4298&(0.88,0.05,0.02,0.06)&0.57&0.21&0.11&&&&&&0.25\\

$\phantom{-}$& $\Sigma_c \bar{D}^*(4462)$&4426&(0.24,0.10,0.47,0.20)&17.53&0.01&0.15&&10.61&1.63&&&5.13\\ [2pt]
\hline
$\phantom{-}$& $\Sigma_c^*\bar{D}^*(4526)$&4509&(0.77,0.10,0.13,0)&1.87&&0.28&0.08&&&&0.43&1.08\\

$\frac{3}{2}$ & $\Sigma_c^*\bar{D}(4386)$ &4376&(0.95,0.05,0,0)&1.06&&0.35&&&&&&0.71\\ [2pt]

$\phantom{-}$& $\Sigma_c\bar{D}^*(4462)$ &4457& (0.95,0.02,0.01,0.02) &1.00&&0.09&0.61&&&&&0.31\\
 \end{tabular}
 \end{ruledtabular}
 \end{table*}

It is found in Table \ref{tab:mol1} that six mass eigenstates of isospin 1/2 below the mass thresold. One may name them $X(4298)$, $X(4426)$, $X(4444)$, $X(4457)$, $X(4378)$ and $X(4509)$. Except for the two spin 1/2 states of 4444 and 4426 MeV, others are dominated by hadronic molecules since the main contribution of their wave functions is from the molecular structure. 

To check the stability of $X$ states as hadronic molecules, we vary the mass of the compact pentaquark states. The dependence of the mass of the $X$ states on the pentaquark mass change, $\Delta E_{Penta}$ is shown in Fig. \ref{runmass}, where the mass of all the pure compact pentaquark states in Table \ref{tab:mol1} is changed the same, but the model parameters including the constituent quark masses are not changed. 
It is found that for X(4298), X(4457), X(4378) and X(4509) states, the masses are very stable with the change of compact pentaquark masses. The X(4444) and X(4426) are sensitive to the mass change $\Delta E_{Penta}$ since they have larger components of compact pentaquarks. When the mass of all the pentaquark states coupled with the $\Sigma_c\bar{D}^{*}$ system is increased by over 25 MeV, $X(4444)$ goes above the $\Sigma_c\bar{D}^*$ mass threshold. Therefore, the $X(4444)$ is unlikely to be a hadronic molecule dominant state. The hadronic molecular component likely plays an important role in $X(4426)$ since the $X(4426)$ is still well below the $\Sigma_c\bar{D}^{*}$ mass threshold even when the mass of the coupled compact pentaquark components is increased by 50 MeV. 

Considering only the mass closeness, one may propose a very tentative assignment: X(4298) to $P_c(4312)^+$ being a $\Sigma_c\bar{D}$ hadronic molecule with $J^P =1/2^-$, $X(4378)$ to $P_c(4380)^+$ being  a $\Sigma_c^*\bar{D}$ hadronic molecule with $J^P=3/2^-$, $X(4457)$ to $P_c(4457)^+$ being a $\Sigma_c\bar{D}^*$ hadronic molecule with quantum numbers $3/2^-$, and $X(4444)$ and/or $X(4426)$ to $P_c(4440)^+$ being a compact pentaquark dominant state with a considerable $\Sigma_c\bar{D}^*$ component. 

Hadronic molecular states with isospin $I=3/2$ are also predicted, as shown in Table \ref{tab:i32}. We do not discuss them here since there is no any experimental data available in the market.  

\section{\label{sec4}Decay widths}

\begin{figure}[!t]
\centering
\begin{subfigure}[b]{0.4\textwidth}
\centering
\includegraphics[scale=0.4]{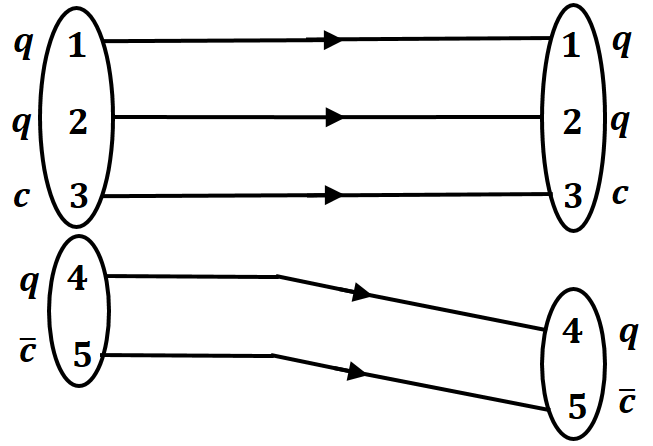}
\caption{\label{da}}
\end{subfigure}
\quad
\begin{subfigure}[b]{0.4\textwidth}
\centering
\includegraphics[scale=0.4]{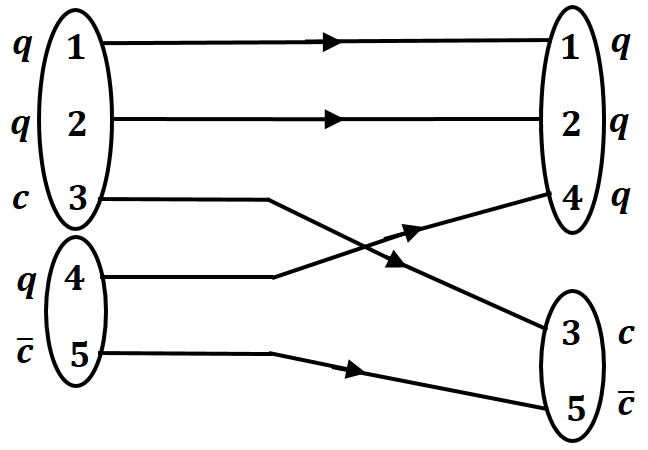}
\caption{\label{db}}
\end{subfigure}
\caption{Quark rearrange diagrams for $P_c$ in (a) the open-charm decay modes and (b) the hidden-charm decay modes.\label{diagram}}
\end{figure}

We study in this section the two-body strong decay property of the bound states, $X(4298)$, $X(4426)$, $X(4457)$, $X(4378)$ and $X(4509)$ in the rearrangement mechanism. The transition amplitude may be defined, 
\begin{eqnarray}\label{binding}
T&=& T^{CSF}\langle \psi_{f} |\hat{O} |P_c \rangle
\end{eqnarray}
with $\hat O$ taking the form,
\begin{eqnarray}
	\label{delta1}
	\hat O_{d}=&  \delta^3(\vec{p}_1-\vec{p'_1})\delta^3(\vec{p}_2-\vec{p'_2})\delta^3(\vec{p}_3-\vec{p'_3})\delta^3(\vec{p}_4-\vec{p'_4}) \nonumber\\&\delta^3(\vec{p}_5-\vec{p'_5}),\\
	\label{delta2}
	\hat O_{c} =&  \delta^3(\vec{p}_1-\vec{p'_1})\delta^3(\vec{p}_2-\vec{p'_2})\delta^3(\vec{p}_3-\vec{p'_4})\delta^3(\vec{p}_4-\vec{p'_3})\nonumber\\&\delta^3(\vec{p}_5-\vec{p'_5}).
\end{eqnarray}
for the processes in Fig.~\ref{diagram}.
$|P_c\rangle$ are the spatial wave function of the $P_c$ states, $X(4298)$, $X(4426)$, $X(4457)$, $X(4378)$ and $X(4509)$, which are derived by solving the coupled Schr\"odinger equations in the mixing picture of hadronic molecules and compact pentaquark states. $\psi_{f}$ are the spatial wave functions of the final states of all possible two-body strong decay channels: $NJ/\psi$, $N\eta_c$, $\Sigma_c^*\bar{D}^*$, $\Sigma_c\bar{D}^*$, $\Lambda^+_c\bar{D}^*$, $\Sigma_c^*\bar{D}$, $\Sigma_c\bar{D}$ and $\Lambda^+_c\bar{D}$. The wave functions of the baryons and mesons of the final states are taken from  the works \cite{Kai2020PRD,Zhao2021PRD} where their mass spectra are fitted. 
$T^{CSF}$ are the color-spin-flavor transition factors of the two-body strong decay channels, as obtained in Ref. \cite{Ruangyoo_2022}. 

The partial decay width can be evaluated in the non-relativistic approximation \cite{book1,ajayut},
\begin{eqnarray}\label{transition2}
\Gamma_{P_c \rightarrow BM} =  \frac{2 \pi E_1 E_2}{ M} \frac{k}{2 S_i +1} \sum_{m_i,m_j} |T(k)| ^2,
\end{eqnarray}
where $S_i$, $m_i$, and $M$ are the spin, spin projection quantum number, and mass of the initial pentaquark states, respectively. $E_1$ and $E_2$ are the energies of the baryon and meson in the final states. $T(k)$ is the transition amplitude derived in Eq. \ref{binding}. The summation is over the spins of the initial and final states. $k$ is the momentum of baryons or mesons at the rest frame of the initial state. 

The partial and total decay widths of $X(4298)$, $X(4426)$, $X(4457)$, $X(4378)$ and $X(4509)$ are calculated in Eq.~(\ref{transition2}) and listed in Table \ref{molepartial1}. The total decay width of the state $X(4457)$ is set to be 1, and the other partial decay widths are all normalized according to this state. It is found in Table \ref{molepartial1} that the decay widths of X(4298), X(4378), X(4457) and X(4509) are in the same order while X(4426) has a much bigger decay width. The results are consistent with the mass spectrum calculations in Section II, where X(4298), X(4378), X(4457) and X(4509) are found dominantly hadronic molecules while X(4426) has considerable both the molecular and compact pentaquark components. 

%  \begin{table}[b]
%  \setlength{\tabcolsep}{2pt}
% \centering
% \caption{The dependence of the ratios of total decay widths between $P_c$ states on the varied pentaquark mass.}\label{ratio1}
% \begin{ruledtabular}
% \begin{tabular}{clll}
%  $P_c$&  $\frac{\Gamma(P_c(4312))}{\Gamma(P_c(4457))}$  & $ \frac{\Gamma(P_c(4380))}{\Gamma(P_c(4457))}$ & $\frac{\Gamma(P_c(4440))}{\Gamma(P_c(4457))}$   \\
%    \hline
% $\phantom{-}$ & $1.5$ & $32$ & $3.2$ \\
%  \hline
% $\Delta E_{Penta}$ &  $\frac{\Gamma(X(4298))}{\Gamma(X(4457))}$ & $ \frac{\Gamma(X(4376))}{\Gamma(X(4457))}$ & $\frac{\Gamma(X(4426))}{\Gamma(X(4457))}$  \\
% \hline
%% -50   & 0.12 & 52& 52 & 11 \\
% -25   & 0.14 &0.617 &9.8\\
% 0   & 0.57 &1.06 &17.5\\
%%  15   & 1.6 &32.1&20.0&2.54\\
% +25   & 2.5 &3.5 &33.1\\
% +50  & 3.9 &4.7&21.3\\
% \end{tabular}
%  \end{ruledtabular}
% \end{table}

\section{\label{sec5}Discussion and Summary}
We have calculated the mass spectrum and strong decay widths of the ground hidden-charm pentaquark states in the mixing picture of hadronic molecules and compact pentaquark states coupled by the one-gluon exchange hyperfine interaction. The work predicts that four $I=1/2$ states, X(4298), X(4378), X(4457) and X(4509) are dominantly hadronic molecules, and one $I=1/2$ state, X(4426) has sizable components of both the compact pentaquark states and hadronic molecules. The X(4298), X(4378), X(4457) and X(4509) have much smaller decay widths than the X(4426). 

Considering our previous work \cite{Ruangyoo_2022} and the work of others \cite{Gian2019,Zhupenta2019,RZhu2019} which predict that the mass of compact charmonium-like pentaquarks is well above the $P_c (4312)$, we may assign the $X(4298)$ to be the $P_c (4312)$, and accordingly $X(4457)$ to be the $P_c (4457)$. 

The work predicts a $J^P=3/2^-$ $\Sigma_c^*\bar{D}$ molecular state, X(4378) which has a decay width in the same order as X(4298) and X(4457). The present result is consistent with Ref. \cite{PhysRevLett.124.072001} where a narrow $P_c$(4380) of $3/2^-$ is predicted in the coupled channel analysis with one-pion exchange and heavy quark spin symmetry. 
The X(4378) in the work can not be assigned to the $P_c$(4380) reported by LHCb if one has assigned the $X(4298)$ to be the $P_c (4312)$. The existence of the wide pentaquark state, $P_c$(4380) still awaits for the verification of larger dataset in the future. 

We suggest that charmonium-like pentaquarks may be searched in the $\eta_c p$ and $\Lambda_c \bar D^{(*)}$ channels in future experiments.

\begin{acknowledgments}
This work is supported by Suranaree University of Technology (SUT) and National Cheng Kung University (NCKU). K.X. is also supported by (i) Thailand Science Research and Innovation (TSRI), and (ii) National Science, Research and Innovation Fund (NSRF) (project code 90464 for Full-Time61/01/2021) .K.P., W.R., and C.C.C. acknowledge support from SUT and NCKU. A.L., and Y.Y. acknowledge support from the NSRF via the Program Management Unit for Human Resources and Institutional Development, Research and Innovation [Grant Number. B05F640055]. 

\end{acknowledgments}
\bibliographystyle{unsrt}

%\bibliography{bibpcmix}% Produces the bibliography via BibTeX.

\end{document}